\documentclass[%
 showpacs,
 showkeys,
 amsmath,amssymb,
]{revtex4-1}

\usepackage{graphicx}% Include figure files
\usepackage{dcolumn}% Align table columns on decimal point
\usepackage{bm}% bold math
\usepackage{CJK}

\usepackage{color}
\usepackage{amssymb}
\usepackage{amsfonts}

\usepackage[colorlinks,urlcolor=blue,linkcolor=blue,citecolor=blue,anchorcolor=blue]{hyperref}

\begin{document}

%\begin{CJK}{GBK}{song}

\preprint{APS/123-QED}

\title{$\mathcal{PT}$ symmetry in a fractional Schr\"odinger equation}% Force line breaks with \\

\author{Yiqi Zhang$^1$}
\email{zhangyiqi@mail.xjtu.edu.cn}
\author{Hua Zhong$^1$}
\author{Milivoj R. Beli\'c$^{2}$}
\author{Yi Zhu$^3$}
\author{Weiping Zhong$^4$}
\author{Yanpeng Zhang$^{1}$}
\author{Demetrios N. Christodoulides$^5$}
\author{Min Xiao$^{6,7}$}
%\email{mxiao@uark.edu}
\affiliation{%
 $^1$Key Laboratory for Physical Electronics and Devices of the Ministry of Education \& Shaanxi Key Lab of Information Photonic Technique,
Xi'an Jiaotong University, Xi'an 710049, China \\
$^2$Science Program, Texas A\&M University at Qatar, P.O. Box 23874 Doha, Qatar \\
$^3$Zhou Pei-Yuan Center for Applied Mathematics, Tsinghua University, Beijing 100084, China \\
$^4$Department of Electronic and Information Engineering, Shunde Polytechnic, Shunde 528300, China\\
$^5$CREOL, College of Optics and Photonics, University of Central Florida, Orlando, Florida 32816-2700, USA\\
$^6$Department of Physics, University of Arkansas, Fayetteville, Arkansas 72701, USA \\
$^7$National Laboratory of Solid State Microstructures and School of Physics, Nanjing University, Nanjing 210093, China
}%

\date{\today}% It is always \today, today,

\begin{abstract}
  \noindent
  We investigate the fractional Schr\"odinger equation with a periodic $\mathcal{PT}$-symmetric potential.
  In the inverse space, the problem transfers into a first-order nonlocal frequency-delay partial differential equation.
  We show that at a critical point,
  the band structure becomes linear and symmetric in the one-dimensional case,
  which results in a nondiffracting propagation and conical diffraction of input beams.
  If only one channel in the periodic potential is excited,
  adjacent channels become uniformly excited along the propagation direction,
  which can be used to generate laser beams of high power and narrow width.
  In the two-dimensional case, there appears conical diffraction that depends on
  the competition between the fractional Laplacian operator and the $\mathcal{PT}$-symmetric potential.
  This investigation may find applications in novel on-chip optical devices.
\end{abstract}

%\pacs{03.65.Ge, 03.65.Sq, 42.25.Gy}% PACS, the Physics and Astronomy
%Use showkeys class option if keyword
\maketitle

%\tableofcontents
%
\section{Introduction}

It is well known that a non-Hermitian Hamiltonian may possess an entirely real eigenvalue spectrum if it is $\mathcal{PT}$-symmetric \cite{bender.prl.80.5243.1998}.
$\mathcal{PT}$ symmetry implies that the eigenfunctions of a Hamiltonian are also eigenfunctions of
the parity-time operator,
that is $H \hat{P}\hat{T} = \hat{P}\hat{T} H$.
The action of the parity operator $\hat{P}$ is defined by $\hat{p} \rightarrow -\hat{p},~\hat{x} \rightarrow -\hat{x}$,
whereas that of the time operator $\hat{T}$ by $\hat{p} \rightarrow -\hat{p},~\hat{x} \rightarrow \hat{x},~i \rightarrow -i$,
where $\hat{p}$ and $\hat{x}$ represent the momentum and position operators.
From this point of view, a $\mathcal{PT}$-symmetric Hamiltonian requires
$V(x)=V^*(-x)$, which indicates that the real part of the complex potential should be an even function of the position and the imaginary part should be an odd \cite{bender.rpp.70.947.2007}.
Such potentials are easily realized in optics.
There, the real part of the complex potential corresponds to an even change in the refractive index,
whereas the imaginary part indicates alternatively changing loss and gain with equal amplitudes \cite{ganainy.ol.32.2632.2007}.

Thus far, it has been theoretically and experimentally established that light possesses unique characteristics
in both linear \cite{makris.prl.100.103904.2008,guo.prl.103.093902.2009,longhi.prl.103.123601.2009,makris.pra.81.063807.2010,ruter.np.6.192.2010,hang.prl.110.083604.2013}
and nonlinear \cite{musslimani.prl.100.030402.2008,zezyulin.prl.108.213906.2012,lumer.prl.111.263901.2013,kartashov.prl.115.193902.2015,makris.nc.6.7257.2015,wimmer.nc.6.7782.2015}
$\mathcal{PT}$-symmetric systems.
In particular, the spectra of $\mathcal{PT}$-symmetric systems possess a critical point, above which the eigenvalues of the system become complex and therefore the $\mathcal{PT}$ symmetry is broken.
It is also known that in topological photonics \cite{rechtsman.nature.496.196.2013,plotnik.nm.13.57.2014,zhang.lpr.9.331.2015,song.nc.6.6272.2015},
Dirac cones in the energy zones can be deformed by the $\mathcal{PT}$ symmetry, which results in the complex eigenvalues \cite{szameit.pra.84.021806.2011,zhen.nature.525.354.2015} becoming distributed in a circular fashion.
Based on these unique properties, enabled by $\mathcal{PT}$ symmetry, some novel on-chip optical devices such as optical isolators \cite{regensburger.nature.488.167.2012,feng.nm.12.108.2013,peng.np.10.394.2014,chang.np.8.524.2014}
and coherent perfect absorbers \cite{wan.science.331.889.2011,chong.prl.105.053901.2010,sun.prl.112.143903.2014} have been proposed.
Very recently it was pointed out how a complex potential can be used to illustrate specific features of disordered systems,
in particular non-Hermitian photonic lattices \cite{longhi.sr.5.13376.2015} and optical waveguide arrays \cite{kartashov.lpr.10.100.2016}.

On the other hand, research into fractional Schr\"odinger equation (FSE) \cite{laskin.pre.66.056108.2002}
-- a generalization of the standard Schr\"odinger equation (SE) that includes fractional derivatives --
lead to an interesting extension of quantum mechanics that included new insights into the fractional field theory and the behavior of particles with fractional spin \cite{herrmann.book.2011}.
Another interesting example of space-fractional quantum mechanics is a condensed-matter realization of L\'evy crystals \cite{stickler.pre.88.012120.2013}.
The FSE was introduced into optics in 2015, and both the steady behavior \cite{longhi.ol.36.2883.2015}
and propagation dynamics \cite{zhang.prl.115.180403.2015} of wavepackets in a harmonic potential were investigated.
Different from the phenomena observed in a regular SE, those found in FSE are truly intriguing,
such as the zigzag propagation of light in a parabolic potential.
Most of the problems related to FSE are still open, for the simple reason -- the rigorous mathematical foundation of many facets of fractional calculus is still missing.
A quintessential example is the one dealing with light manipulation and control with the help of a $\mathcal{PT}$-symmetric potential.
Indeed, the modulation of light in such a system represents a fresh problem of practical interest that deserves deeper exploration.

In this paper, we investigate the dynamics of waves in the FSE with a $\mathcal{PT}$-symmetric potential.
We find that similar phenomena to the ones observed in the regular SE can be also observed in this model.
Still, major differences abound.
The band structure of such a model is different from that of a regular SE, especially near the critical point,
which becomes kink-like in the one-dimensional (1D) case and cone-like in the two-dimensional (2D) case.
The linear band structure indicates a nondiffracting propagation in the 1D case.
However, whether conical diffraction in the 2D case can be realized or not,
depends on the strength of the $\mathcal{PT}$-symmetric potential.

\section{One-dimensional case}

The FSE we are interested in can be written as
\begin{equation}\label{60eq1}
  i\frac{\partial \psi}{\partial z} + \left[ -\left(-\frac{\partial^2}{\partial x^2}\right)^{\alpha/2} + V(x) \right] \psi = 0,
\end{equation}
where $\alpha$ is the L\'{e}vy index ($1<\alpha\le2$) and $V(x)$ is the periodic $\mathcal{PT}$-symmetric potential.
As a versatile model, we pick the simple periodic lattice potential of the form $V(x)= A [\cos^2(x)+iV_0\sin(2x)]$
with the amplitude $A=4$ and $V_0$ a parameter to be varied.
Clearly, the period of the potential is $D=\pi$.
When the L\'evy index is $\alpha=2$, one recovers the usual SE.
Here, we explore the opposite region, $\alpha$ close to 1 from the positive side \cite{zhang.prl.115.180403.2015}.

The solution of Eq. (\ref{60eq1}) can naturally be written in the form
$\phi_n(x,k)\exp[i\beta_n(k) z]$, in which $\phi_n(x,k)$ is the Bloch mode and $\beta_n(k)$ is the propagation constant.
Plugging this ansatz into Eq. (\ref{60eq1}), one obtains
\begin{equation}\label{60eq2}
  -\beta \phi + \left[ -\left(-\frac{\partial^2}{\partial x^2}\right)^{\alpha/2} + V(x) \right] \phi = 0.
\end{equation}
According to the Floquet-Bloch theorem, $\phi(x)$  can be written as $\phi(x) = w_k(x) \exp(i k x)$, where $w_k(x) = w_k(x+D)$ is spatially periodic.
One can expand $w_k(x)$ and the potential in series of plane-waves, $w_k(x) = \sum_n c_n \exp(iK_n x)$, with $ K_n  = {2 \pi n}/{D}$
and $V(x) = \sum_m P_m \exp(iK_m x)$, where $P_m = \int_D V(x) \exp(-i K_m x) dx /D$.
Plugging these series into Eq. (\ref{60eq2}), one obtains
\begin{align*}
\sum_n \left[ -\beta - |k+K_n|^\alpha \right]  c_n \exp[i(k+K_n)x]
 + \sum_{m,n} P_m c_n \exp[i(k+K_n+K_m)x] = 0.
\end{align*}
Multiplying the above equation by $\exp[-i(k+K_q)x] $ and integrating over $x\in(-\infty,+\infty) $, one ends up with
\[-|k+K_q|^\alpha c_q + \sum_m P_m c_{q-m} = \beta c_q,\]
which is an eigenvalue problem in matrix form.
As a result, the band structure corresponding to Eq. (\ref{60eq2}) can be obtained for certain $\alpha$.

\begin{figure}[htbp]
	\centering
	\includegraphics[width=0.5\columnwidth]{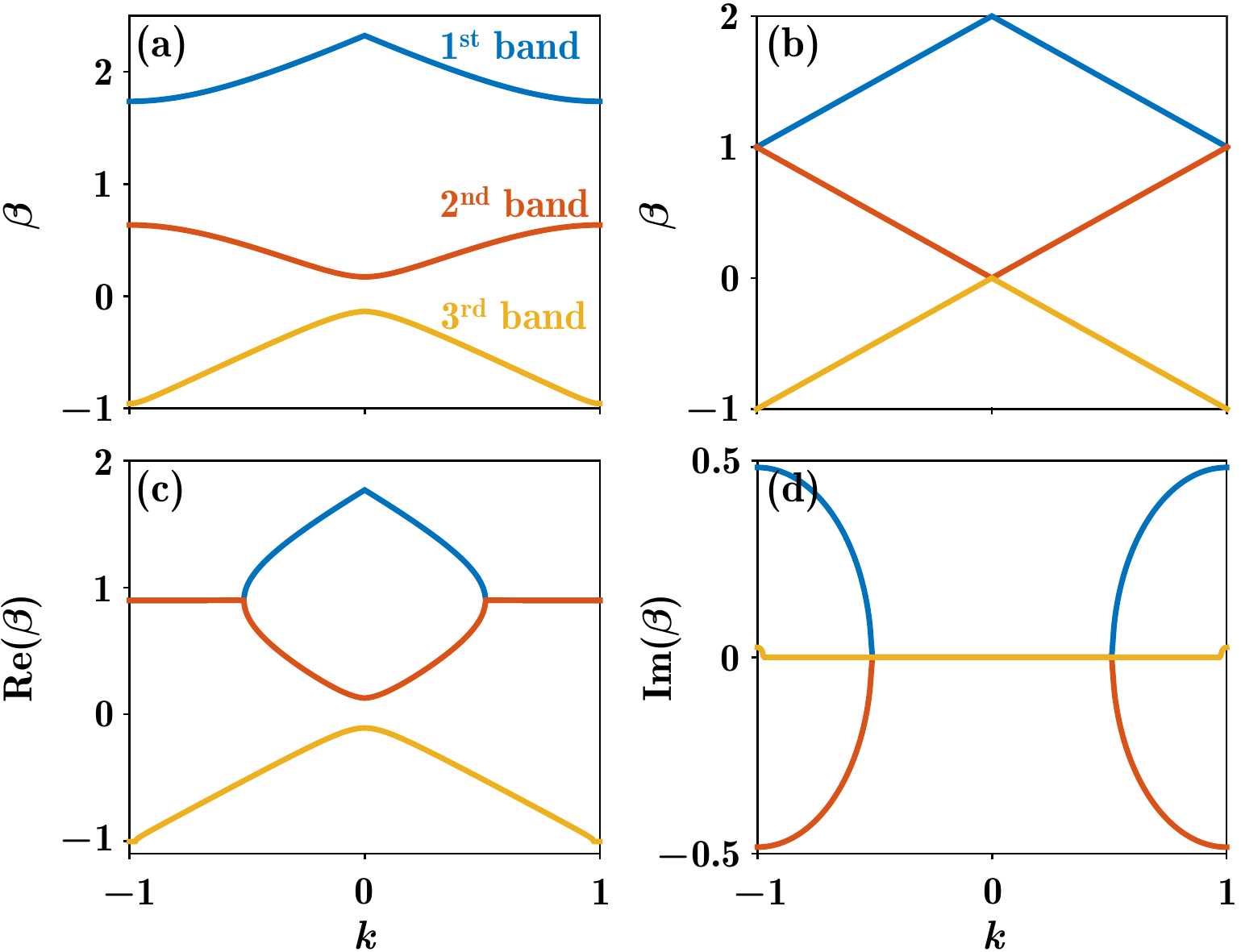}\\
	\caption{(Color online) Photonic band structure of Eq. (\ref{60eq2}) for (a) $V_0=0.4$ and (b) $0.5$.
    (c) and (d) Real and imaginary parts of the band structure for $V_0=0.55$, respectively.}
	\label{60fig1}
\end{figure}

The photonic band structure in the first Brillouin zone, corresponding to different values of $V_0$, is displayed in Fig. \ref{60fig1}.
Clearly, the threshold value of $V_0$ is $V_0^{\rm th}=0.5$,
below which the band structure is entirely real, as shown in Fig. \ref{60fig1}(a).
However, at the threshold value, as shown in Fig. \ref{60fig1}(b), the band gap disappears and the two bands connect with each other at the edges of the Brillouin zone.
Further increasing the value of $V_0$, the two bands begin to merge with each other and become a double-valued band.
In Figs. \ref{60fig1}(c) and \ref{60fig1}(d), the real and imaginary parts of the band structure are displayed.
One can see that the imaginary eigenvalues in Fig. \ref{60fig1}(d) are all zero in the closed region of Fig. \ref{60fig1}(c),
while outside the region, the eigenvalues are complex.
More numerical simulations demonstrate that the closed region in Fig. \ref{60fig1}(c) shrinks gradually and ultimately disappears with increasing $V_0$.
The phenomena described are quite similar to those reported previously \cite{makris.prl.100.103904.2008,makris.pra.81.063807.2010},
but there are also stark differences -- most notably, around $k=0$ the bands are symmetric and almost linear.
Especially, the bands are completely symmetric and linear at $V_0=V_0^{\rm th}$, as displayed in Fig. \ref{60fig1}(b).
Linear bands mean that a beam propagating in this lattice is diffraction-free, because the second-order derivative of the band is 0.

We should note that the band structure at the critical point $V_0=0.5$ can also be derived analytically,
based on the method previously applied to the regular Schr\"odinger equation ($\alpha=2$) \cite{longhi.pra.81.022102.2010}.
Rather generally, at $ V_0=0.5$ the band structure of the fractional Schr\"odinger equation is the same as
the one of the potential-free particle with fractional kinetic energy operator, i.e. of the type $\sim|k|^\alpha$ \cite{zhang.arxiv.2015}.
For the limiting cases, at $\alpha=2$ one has the usual parabolic dispersion relation,
and at $\alpha=1$ one finds the linear dispersion relation leading to conical diffraction.

\begin{figure}[htbp]
	\centering
	\includegraphics[width=0.5\columnwidth]{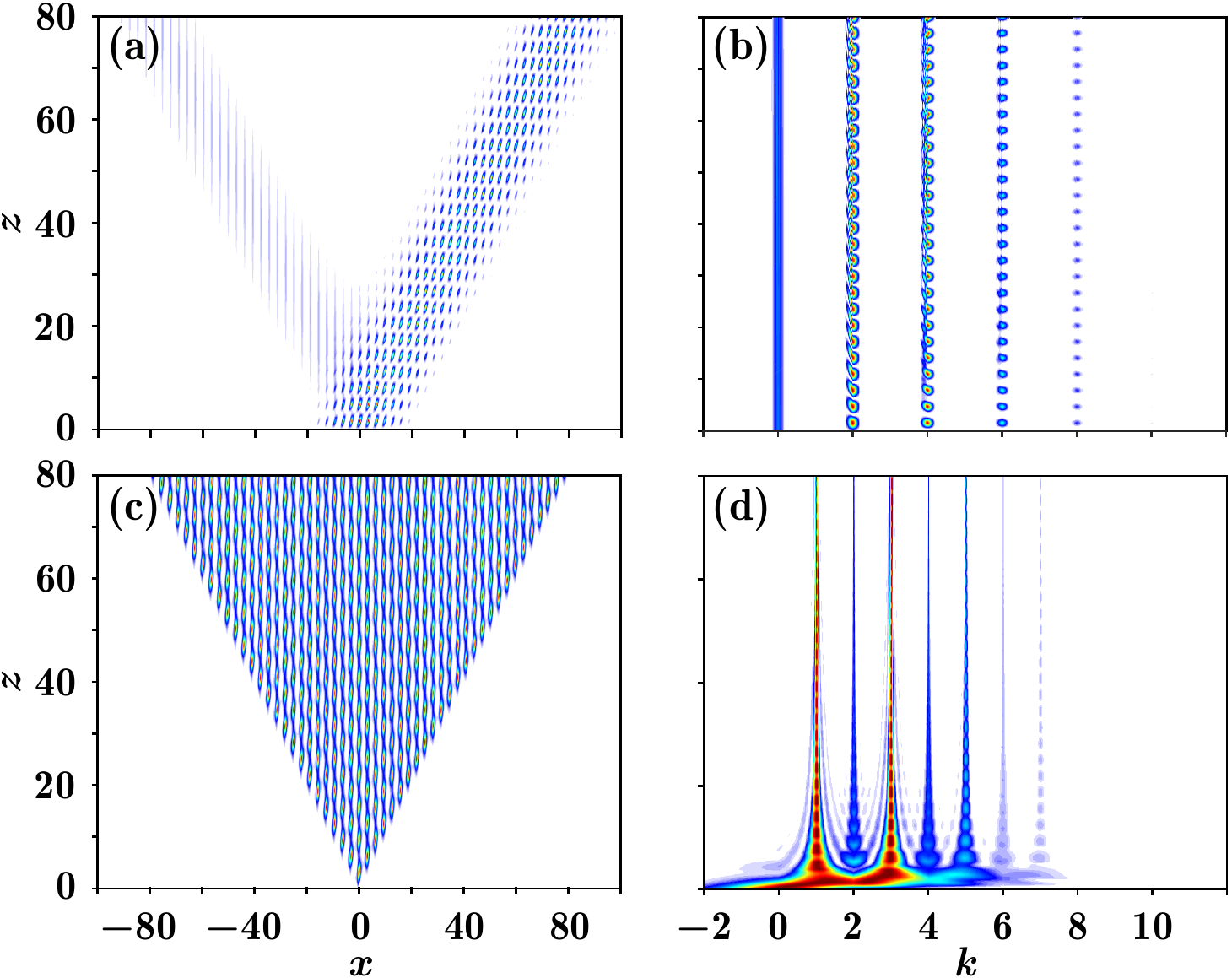}\\
	\caption{(Color online) (a) Propagation of a wide Gaussian beam $\psi(x)=\exp(-x^2/100)$ in the $\mathcal{PT}$-symmetric lattice with $V_0=0.5$.
    (b) The corresponding propagation in the inverse space.
    (c) and (d) Setup is as in (a) and (b), but with a narrow Gaussian beam $\psi(x)=\exp(-x^2)$ that excites only one channel.}
	\label{60fig2}
\end{figure}

We now focus on the propagation of light at the transition point.
It has to be done numerically, and we do it by utilizing the split-step fast Fourier transform method of second order, as displayed in Fig. \ref{60fig2}.
Figure \ref{60fig2}(a) shows the case of many waveguiding channels excited by a wide Gaussian beam;
the beam indeed propagates without diffraction except for the initial splitting, which is a 1D-equivalent of conical diffraction.
Note the preferential propagation to one side, which can be explained by the projection coefficient method of \cite{makris.prl.100.103904.2008,makris.pra.81.063807.2010}.
Unidirectional propagation can be observed if the beam is launched with an appropriate incident angle.
Conical diffraction results from the symmetric linear diffraction relation,
which is similar to those discussed in topological photonics \cite{zeuner.prl.109.023602.2012,dubvcek.njp.17.125002.2015}.

To better understand the propagation, we consider Eq. (\ref{60eq1}) in the inverse space, which can be written as
\begin{align}\label{60eq3}
  i\frac{\partial \hat\psi(k,z)}{\partial z} + c_0\hat\psi(k,z) 
   + c_{-2} \hat\psi(k-2,z) + c_{+2} \hat\psi(k+2,z) = 0,
\end{align}
where $\hat\psi$ is the Fourier transform of $\psi$, $c_0=A/2- |k|$, $c_{-2}=A\pi({1}/{2}+V_0)$, and $c_{+2}=A\pi({1}/{2}-V_0)$.
Clearly, Eq. (\ref{60eq3}) is a first-order nonlocal frequency-delay partial differential equation
that couples the wavefunction at different $k$ values,
similar to the coupled discrete photonic waveguides \cite{garanovich.pr.518.1.2012}.
Such an infinitely-dimensional dynamical system is quite typical of FSE,
in which the coupling strengths for the cases $k\leftrightsquigarrow k-2$ on the $k>0$ side
and $k\leftrightsquigarrow k+2$ on the $k<0$ side are different.
This explains the observed fact that light is preferentially skewed to one side during propagation in a $\mathcal{PT}$-symmetric system.
At the critical point $V_0=1/2$, the coupling exists only for the $k\leftrightsquigarrow k-2$ case.
Figure \ref{60fig2}(b), which is the propagation in the inverse space that corresponds to Fig. \ref{60fig2}(a),
demonstrates that such a coupling only acts on the $k>0$ side,
where the new modes in the inverse space are discretely excited during propagation.
It is worth mentioning that linear dispersion relations and conical diffraction in a $\mathcal{PT}$-symmetric waveguide array
with the regular diffraction band structure can also be observed by introducing longitudinal gain/loss modulation \cite{della.87.022119.2013}.

We now turn to the case with only one channel initially excited, as depicted in Figs. \ref{60fig2}(c) and \ref{60fig2}(d).
In Fig. \ref{60fig2}(c), one finds that the traditional discrete diffraction \cite{lederer.pr.463.1.2008,longhi.lpr.3.243.2009} disappears,
and instead the channels are excited gradually and uniformly to the left and right,
which results in an increasing power of the beam.
Note that such behavior was explained before on the basis of spectral singularities of the Schr\"odinger equation at the critical point \cite{longhi.pra.81.022102.2010}.
Even though the power is not conserved during propagation, the ``quasipower'' $Q=\int^{+\infty}_{-\infty} \psi(x) \psi^*(-x) dx$ is.
From Fig. \ref{60fig2}(d), which represents the corresponding propagation in the inverse space,
the coupling appears in the $k>0$ region,
which reflects the skewness of the propagation in each channel, noted in Fig. \ref{60fig2}(c).
We should stress that the discrete-like diffraction in Fig. \ref{60fig2}(c) is just
the \textit{conical diffraction} propagation of the case with only one channel excited.

One may notice that the propagation in Fig. \ref{60fig2}(d) seemingly conflicts with the rule unveiled by Eq. (\ref{60eq3}).
However, numerical simulations indicate that the beam stripes at $k=1,\,3,\,5,\cdots$ come from those at $k=2,\,4,\,6,\cdots$.
The reason is that the narrow beam in the real space possesses a broad width in the inverse space,
which leads to the interference among more and more components as the beam propagates.
Note also the increased intensity of $k=1$ and $k=3$ modes, coming from the gain/loss feature of the complex potential.
The intensity shown in Fig. \ref{60fig2}(d) is normalized, to show the propagation more clearly;
the increasing power during propagation will gradually overwhelm the anterior propagation.
One may conclude that filamentary characteristics displayed in Fig. \ref{60fig2}(d), resulting from Fig. \ref{60fig2}(c),
can be used to generate \textit{high-power narrow-width} laser beams.

\begin{figure}[htbp]
	\centering
	\includegraphics[width=0.5\columnwidth]{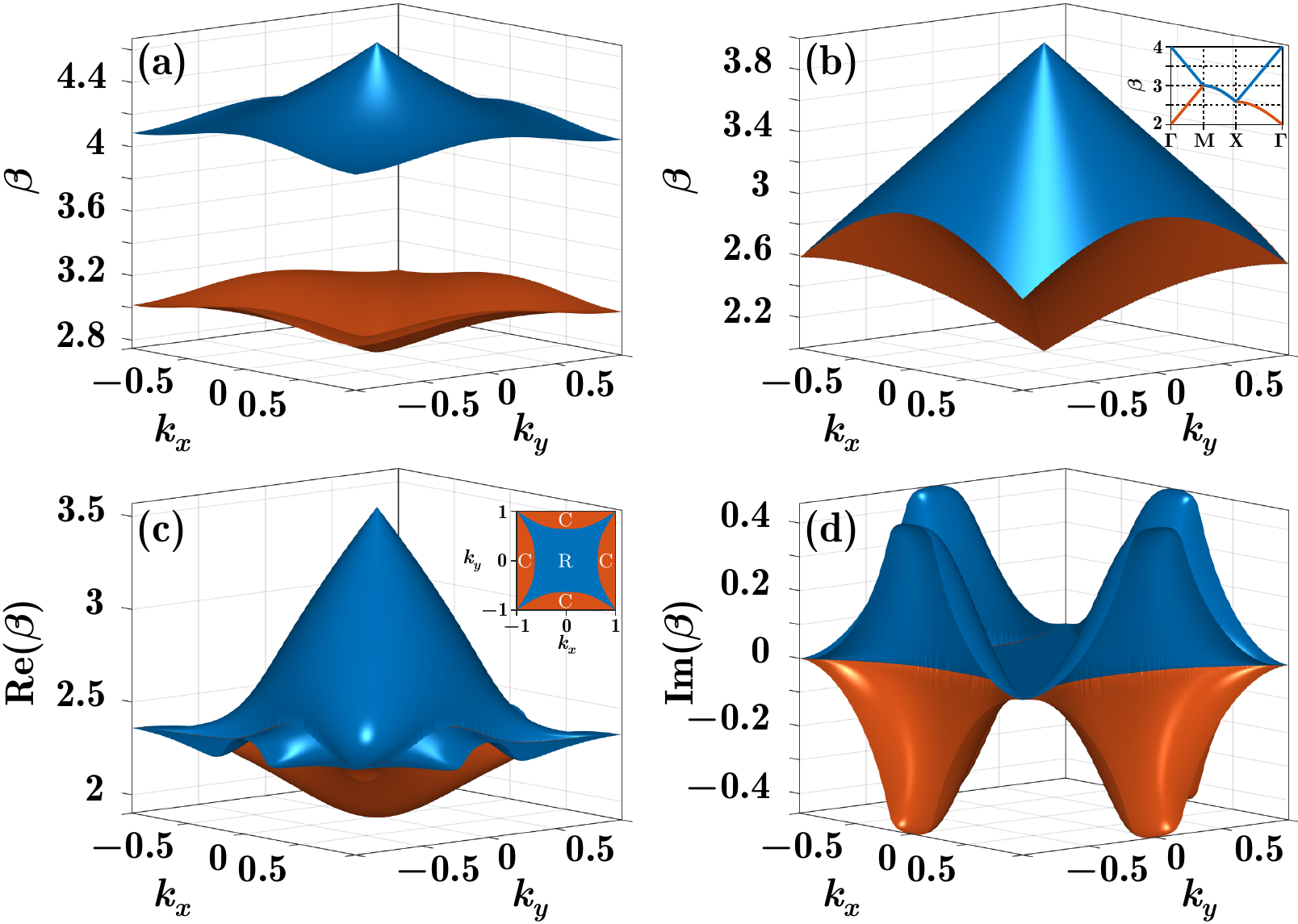}\\
	\caption{(Color online) Two-dimensional band structure of Eq. (\ref{60eq4}).
    Figure setup is as in Fig. \ref{60fig1}.
    Inset in (b) displays the band structure along main symmetry points,
    in which the coordinates are $\Gamma$(0,0), M(1,0), and X(1,1), respectively.
    Inset in (c) shows the phase diagram of the first Brillouin zone,
    in which ``R'' represents the complete real eigenvalue region, and ``C'' represents the complex eigenvalue region.}
	\label{60fig3}
\end{figure}

\section{Two-dimensional case}

We now extend the analysis to FSE in two-dimensions,
which offers even more interesting behavior.
In Cartesian coordinates, the equation can be written as
\begin{equation}\label{60eq4}
  i\frac{\partial \psi}{\partial z} + \left[ -\left(-\frac{\partial^2}{\partial x^2}-\frac{\partial^2}{\partial y^2}\right)^{\alpha/2} + V(x,y) \right] \psi = 0.
\end{equation}
The complex periodic potential is generalized to the appropriate 2D form $V(x,y)=A\{\cos^2(x)+\cos^2(y)+iV_0[\sin(2x)+\sin(2y)]\}$.
The 2D band structure is displayed in Fig. \ref{60fig3}.
Again, the diffraction relation is almost linear at the center of the first Brillouin zone.
Especially, the upper band at the critical point, as shown in Fig. \ref{60fig3}(b), is completely linear.
For the lower band, as exhibited in the inset in Fig. \ref{60fig3}(b), the band is also linear along the $\Gamma$-M direction.
In Figs. \ref{60fig3}(c) and \ref{60fig3}(d), the band structure is for the case above the critical point,
from which one finds that the completely real eigenvalues are surrounded by four symmetric regions of complex eigenvalues,
with the four vertices placed at the corners of the first Brillouin zone, as shown in the inset in Fig. \ref{60fig3}(c).
This is different from the structure displayed in Ref. \cite{makris.prl.100.103904.2008}.
Numerical simulations demonstrate that the region of real eigenvalues is not affected much by the value of $V_0$,
but along the diagonal directions of the first Brillouin zone, the eigenvalues are all real.
In addition, the regions of complex eigenvalues broaden if $V_0$ further increases.

\begin{figure}[htbp]
	\centering
	\includegraphics[width=0.5\columnwidth]{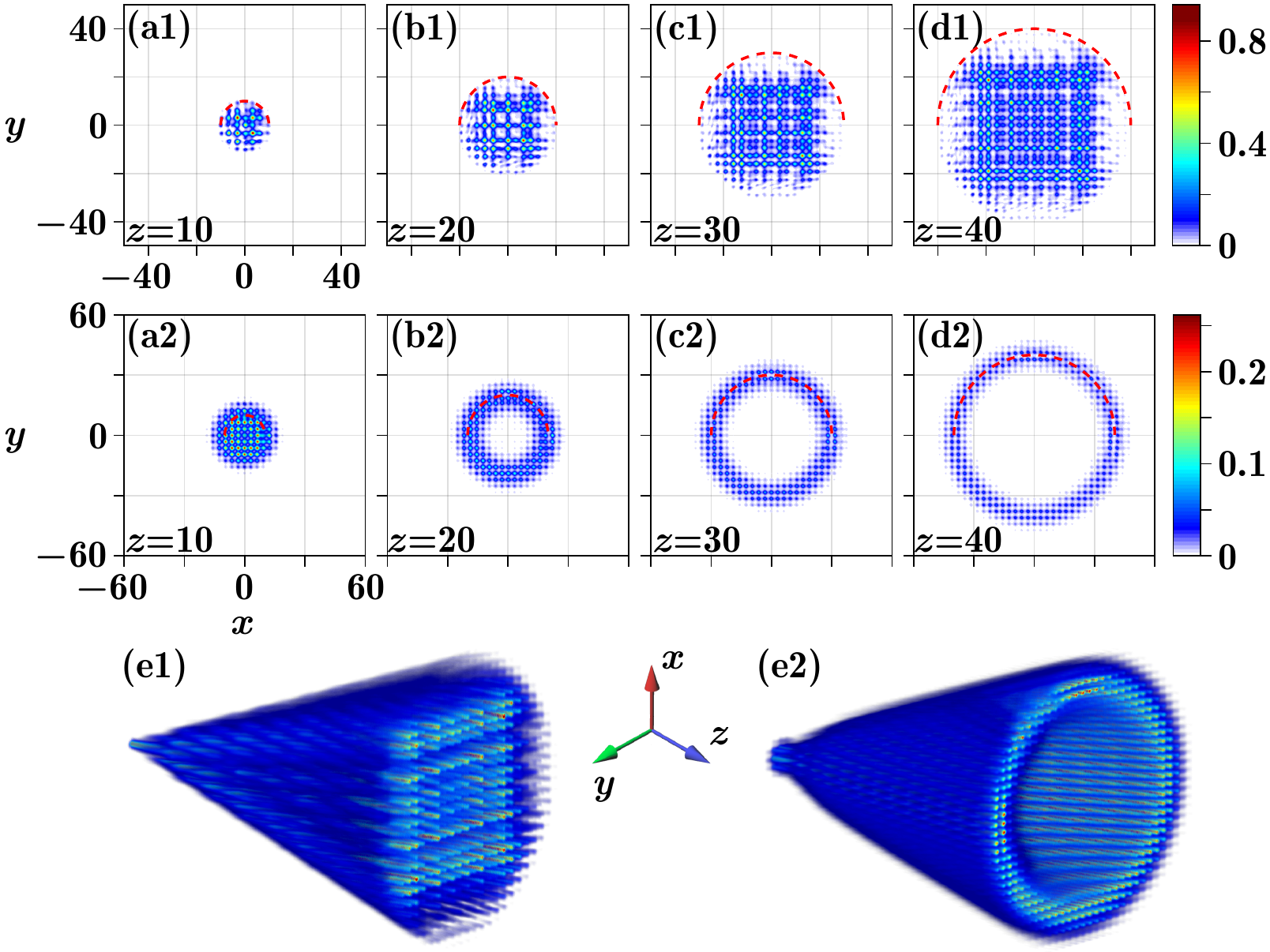}\\
	\caption{(Color online)
    Beam intensities at certain propagation distances, as displayed in each panel.
    The dashed half-circles indicate the theoretical prediction of the conical diffraction.
    (a1)-(d1) The case with only one channel excited.
    (a2)-(d2) The case with many channels excited.
    (e1) and (e2) Panoramic views of the conical diffraction, corresponding to (a1)-(d1) and (a2)-(d2), respectively.}
	\label{60fig4}
\end{figure}

Considering that the upper band in Fig. \ref{60fig3}(b) is completely linear --
which is quite similar to the Dirac cone in topological photonics --
a beam that excites the Floquet-Bloch mode of the upper band will exhibit a 2D conical diffraction \cite{peleg.prl.98.103901.2007,ablowitz.pra.79.053830.2009,song.nc.6.6272.2015} during propagation.
However, such a conical diffraction is affected by the strength of the $\mathcal{PT}$-symmetric potential,
which is a complex square lattice (the first Brillouin zone of which is also a square).

In Figs. \ref{60fig4}(a1)-\ref{60fig4}(d1) and Figs. \ref{60fig4}(a2)-\ref{60fig4}(d2),
we display the beam intensities at certain propagation distances with $A=1$,
corresponding to narrow and wide inputs, respectively.
We have constructed the inputs by multiplying a Gaussian with the Floquet-Bloch mode at $k_x=k_y=0$ \cite{makris.pra.81.063807.2010}.
Figures \ref{60fig4}(e1) and \ref{60fig4}(e2) exhibit the corresponding panoramic view of the propagation.
Clearly, the conical diffraction can be realized in the case of a relatively weak $\mathcal{PT}$-symmetric potential.
According to Fig. \ref{60fig3}(b) and the inset, the group velocity of the conical diffraction should be 1,
since the slope of the linear band structure is 1.
Therefore, the relation between the radius of the cone and the propagation distance should be
$
z=\sqrt{x^2+y^2}.
$
In Figs. \ref{60fig4}(a1)-\ref{60fig4}(d1) and Figs. \ref{60fig4}(a2)-\ref{60fig4}(d2),
the dashed half circles represent the theoretical prediction of the cone size at certain distances,
which completely agrees with the numerical simulations.

One also finds that in Fig. \ref{60fig4}, the beam intensity is symmetric with respect to $y=x$ but asymmetric with respect to $y=-x$;
this is due to the skewness of beam propagation that is caused by the $\mathcal{PT}$-symmetric potential.
If we increase the value of $A$ to 4 (not shown),
the conical diffraction is much inhibited
and the maximum intensity also increases dramatically during propagation,
which means that the $\mathcal{PT}$-symmetric potential plays a major role in the propagation dynamics.
In other words, there is a competition between the fractional Laplacian that leads to the linear band structure in Fig. \ref{60fig3}(b)
(which will induce the conical diffraction \cite{zhang.arxiv.2015} during propagation)
and the $\mathcal{PT}$-symmetric potential (which will induce beam localization).
The main factor that determines the outcome of competition is the strength of the $\mathcal{PT}$-symmetric potential,
which is characterized by the value of $A$.
If the strength of the potential is high enough, the beam will be trapped in the lattice sites to one side and the diffraction will be suppressed \cite{liu.apb.99.727.2010}.

\section{Conclusion and outlook}

In summary, we have investigated the conical diffraction of a light beam in a fractional Schr\"odinger equation with a $\mathcal{PT}$-symmetric potential.
Our investigation not only demonstrates how to obtain beam localization in a $\mathcal{PT}$-symmetric potential without utilizing nonlinearities,
but also connects fractional Laplacian and $\mathcal{PT}$ symmetry,
which indicates that our investigation possesses advantages from both sides.
Thus, it may exhibit a great deal of applicative potential for fabricating on-chip optical devices.

In an outlook, we have to point out that it is still challenging and an open problem
to design a physical realization of the free propagation and conical diffraction in the fractional Schr\"odinger equation with a $\mathcal{PT}$-symmetric potential.
One possible breakout could be based on Eq. (\ref{60eq3}), which is similar to the equation that describes the propagation of light in coupled waveguides.
The complexity here comes from the fact that one has to elaborate on the coupling strength among ``waveguides'',
which indeed is conceivable according to previous literature \cite{garanovich.pr.518.1.2012}.

\section*{Acknowledgements}
This work is supported by National Basic Research Program of China (2012CB921804),
National Natural Science Foundation of China (61308015, 11474228),
Key Scientific and Technological Innovation Team of Shaanxi Province (2014KCT-10),
and Qatar National Research Fund  (NPRP 6-021-1-005).
MRB also acknowledges support by the Al Sraiya Holding Group.

\bibliographystyle{myprx}
\bibliography{my_refs_library}

\end{document}